\documentclass[aps,prl,showpacs,preprintnumbers,twocolumn,superscriptaddress,nofootinbib,floatfix,10pt]{revtex4}
\usepackage{amsfonts,amssymb,stmaryrd,latexsym,amsmath}
\usepackage{graphicx,subfigure}
\usepackage{times}
\usepackage{slashed}

\begin{document}

\preprint{NSF-KITP-16-188}

\title{Effective forces between quantum bound states}

\author{Alexander Rokash}\affiliation{Institut~f\"{u}r~Theoretische~Physik~II,~Ruhr-Universit\"{a}t~Bochum,
D-44870~Bochum,~Germany}

\author{Evgeny Epelbaum}\affiliation{Institut~f\"{u}r~Theoretische~Physik~II,~Ruhr-Universit\"{a}t~Bochum,
D-44870~Bochum,~Germany}
\affiliation{Kavli Institute for Theoretical Physics, University of California,
Santa Barbara, CA 93106-4030, USA}

\author{Hermann~Krebs}
\affiliation{Institut~f\"{u}r~Theoretische~Physik~II,~Ruhr-Universit\"{a}t~Bochum,
D-44870~Bochum,~Germany}
\affiliation{Kavli Institute for Theoretical Physics, University of California,
Santa Barbara, CA 93106-4030, USA}

\author{Dean~Lee}
\affiliation{Department~of~Physics, North~Carolina~State~University, Raleigh, NC~27695, USA}
\affiliation{Kavli Institute for Theoretical Physics, University of California, Santa Barbara, CA 93106-4030, USA}

\begin{abstract}
\noindent
Recent {\it ab initio} lattice studies have found that the interactions between alpha particles ($^4$He nuclei) are  sensitive to seemingly minor details of the nucleon-nucleon force such as interaction locality.  In order to uncover the essential physics of this puzzling phenomenon without unnecessary complications, we study a simple model involving two-component fermions in one spatial dimension. We probe the interaction between two bound dimers for  several different particle-particle interactions and measure an effective potential between the dimers using external point potentials which act as numerical tweezers.  We find that  the strength and range of the local part of the particle-particle interactions
play  a dominant role in shaping the interactions between the dimers and can even determine the overall sign of the effective potential.
            
\end{abstract}

\maketitle
We investigate the connection between microscopic particle interactions and
the cohesion responsible for the formation of complex structures such as nuclei. In atomic physics, the van der Waals force
describes the interaction between neutral spherical atoms due to induced electric dipoles.  The quantum mechanical description was first derived by London using second-order perturbation theory \cite{London:1930}. A similar phenomenon occurs between alpha particles
or $^4$He nuclei, with the two-pion exchange potential playing an important role \cite{Arriola:2007de,Elhatisari:2016owd}. However the physics of this system is significantly more complicated due to the complexity of the nuclear forces and its delicate balance against Pauli repulsion between identical nucleons and the Coulomb interaction.
In Ref.~\cite{Elhatisari:2016owd} a surprising result was found that the local or nearly local part of the nuclear interactions are important in determining the strength of the alpha-alpha interaction. In addition to being an interesting fundamental science question, understanding this phenomenon in more detail may lead to new avenues for improving the order-by-order convergence of {\it ab initio} nuclear structure calculations in medium mass nuclei \cite{Dytrych:2016vjy,Maris:2014hga,Duguet:2016wwr,Stroberg:2016ung,Ruiz:2016gne,Hagen:2016uwj} using chiral effective field theory.  See Ref.~\cite{Epelbaum:2008ga} for a review of chiral effective field theory. 

By a local force we mean that the positions of the particles are left in 
place during the interaction process, while the more general nonlocal force 
may change the relative separation between particles.
We note that there is little difference between an exactly local interaction which keeps the particle positions exactly
in place and a nearly local interaction which changes the particle positions only at length scales small compared to the quantum bound state size. So when we use the term local we mean both cases. In this letter we investigate whether this local dominance of the alpha-alpha interaction could be a general phenomenon. To head towards this goal, we start by examining the simplest possible quantum system in detail to help elucidate the general principles involved.
For this task we study the residual interactions between two bound dimers in one spatial dimension.  Before proceeding we should mention the pioneering work by Combescot and collaborators in understanding the interactions of composite bosons such as excitons composed of elementary fermions 
\cite{Combescot:2005,Combescot:2007,Combescot:2008,Combescot:2009,Combescot:2010}.  Our emphasis though is somewhat different as we focus on the impact of different particle-particle interactions upon the effective dimer-dimer interaction. Besides the main nuclear physics motivation, the results of our lattice study should be useful in predicting the properties of composite bosons on optical lattices for different types of interactions, including single-site and nearest-neighbor interactions.  

\begin{figure}[!ht]
\centering
\caption{Illustration of the external potential acting as numerical tweezers by holding the down-spin particles at separation distance $n$.}
\includegraphics[scale=0.7]{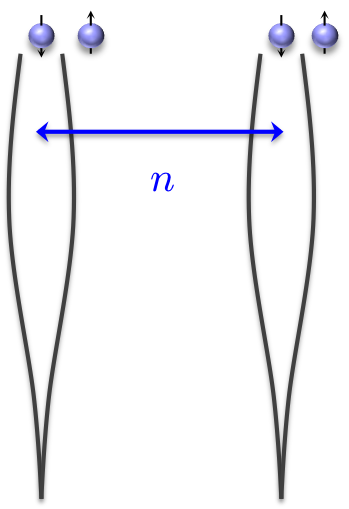}
\label{tweezers_dimers}
\end{figure}

In our calculations we consider two-component fermions, which we label as up and down spins.  As there are different conventions and units used in the  nuclear, atomic, and condensed matter communities, throughout our
analysis we use only dimensionless lattice units. Our 
lattice Hamiltonian has the form,
\begin{align}
& H=t\sum_{n,\sigma=\uparrow,\downarrow}[b_\sigma^{\dagger}(n+1)-b_\sigma^{\dagger}(n)][b_\sigma(n+1)-b_\sigma(n)] \nonumber \\
& + \sum_{n,i_,i'}b_\uparrow^{\dagger}(\tfrac{n+i}{2})b_\downarrow^{\dagger}(\tfrac{n-i}{2})I_{i,i'}b_\downarrow(\tfrac{n-i'}{2})b_\uparrow(\tfrac{n+i'}{2}),
\end{align}
where $n$, $i$, $i'$, $\tfrac{n\pm i}{2}$, and $\tfrac{n\pm i'}{2}$ are restricted to have integer values. We take $t=1/20$ in dimensionless lattice units. In nuclear physics notation where $t$ is usually written as $1/(2m)$, this corresponds with the dimensionless mass parameter $m=10$. We consider five different interactions, and in the following we list all the nonzero terms.  In all
five cases the interaction produces a bound dimer with dimensionless energy $-1/40$. The motivation 
for introducing the various versions of the two-particle interactions will 
be explained below. The first interaction is a point-like interaction with
\begin{equation}
I^{(1)}_{0,0}=c^{(1)},
\end{equation}
where $c^{(1)}=-0.103078.$ The second interaction is a local interaction that has range one lattice unit,\begin{equation}
I^{(2)}_{0,0}=I^{(2)}_{1,1}=I^{(2)}_{-1,-1}=c^{(2)},
\end{equation}
with $c^{(2)}=-0.049851.$ The third interaction is a local interaction that has range two lattice units,
\begin{equation}
I^{(3)}_{0,0}=I^{(3)}_{1,1}=I^{(3)}_{2,2}=I^{(3)}_{-2,-2}=I^{(3)}_{-1,-1}=c^{(3)},
\end{equation}
with $c^{(3)}=-0.038321.$ The fourth interaction is a nonlocal interaction that has range two lattice units and only even-parity interactions,  
\begin{align}
& I^{(4)}_{0,0}=c^{(4)}, \\
& I^{(4)}_{1,1}=I^{(4)}_{2,2}=I^{(4)}_{-2,-2}=I^{(4)}_{-1,-1}=\frac{c^{(4)}}{2}, \\
& I^{(4)}_{1,-1}=I^{(4)}_{2,-2}=I^{(4)}_{-2,2}=I^{(4)}_{-1,1}=\frac{c^{(4)}}{2},
\end{align}
with $c^{(4)}=-0.038321.$ The fifth interaction is another nonlocal interaction with range two lattice units and only even-parity interactions,
\begin{align}
& I^{(5)}_{0,0}=I^{(5)}_{2,2}=I^{(5)}_{-2,-2}=I^{(5)}_{-2,2}=I^{(5)}_{2,-2}=c^{(5)}, \\
& I^{(5)}_{1,1}=I^{(5)}_{-1,-1}= I^{(5)}_{-1,1}=I^{(5)}_{1,-1}=\frac{c^{(5)}}{2},\\
& I^{(5)}_{0,2}=I^{(5)}_{2,0}=I^{(5)}_{-2,0}=I^{(5)}_{0,-2}=-c^{(5)}. 
\end{align}
where $c^{(5)}=-0.052504$. In order to probe the effective forces between two dimers we apply an external potential of the form
\begin{equation}
U_{\rm ext}(n)=
-u\sum_{n'}b_\downarrow^{\dagger}(n')b_\downarrow^{\dagger}(n+n')b_\downarrow(n+n')b_\downarrow(n'),
\label{tweezers}
\end{equation}
with coefficient $u>0$.  We refer to this external potential as numerical tweezers since for large enough $u$ the dimers will be held in place by the point potentials, with the localization length of the tweezer-bound dimer determining the spatial resolution of our external probe. This is illustrated in Fig.~\ref{tweezers_dimers}.

We have made the numerical tweezers translationally invariant
by summing over $n'$ in Eq.~(\ref{tweezers}).  This is convenient for the exact diagonalization calculations we do here.  In other applications such as auxiliary-field Monte Carlo lattice simulations \cite{Elhatisari:2015iga,Lahde:2015ona,Lahde:2013uqa,Epelbaum:2013paa,Elhatisari:2016owd}, a simpler tweezer potential 
with fixed ends is more convenient,  
\begin{equation}
U_{\rm ext}(n)=
-u\,b_\downarrow^{\dagger}(0)b_\downarrow^{\dagger}(n)b_\downarrow(n)b_\downarrow(0).
\label{tweezers2}
\end{equation}
This numerical tweezer technique will be incorporated into future lattice simulations of nuclear structure.

We now compute the ground state energy of the full Hamiltonian with external potential, $H+U_{\rm ext}(n)$, versus separation distance $n$.  This energy, which we call the tweezer effective potential, is measured relative to its limiting value as $n \rightarrow \infty$.  The tweezer effective potential measures the interaction energy between dimers as a function of separation distance.  However there is clearly also some dependence on the manner in which the tweezers are coupled to the dimers.
The tweezers can be viewed as heavy particles which we have coupled to the down-spin particles.  The tweezer effective potential is then the Born-Oppenheimer potential between the heavy particles induced by the dimer-dimer interaction.  

\begin{figure}[!ht]
\centering
\caption{The tweezer effective potential potential versus separation distance for interactions $1$ and $3$ for dimer binding energy
$1/40$ and tweezer couplings $u
= 0.100, 0.125, 0.150$.}
\vspace{0.5cm}
\includegraphics[width=8cm]{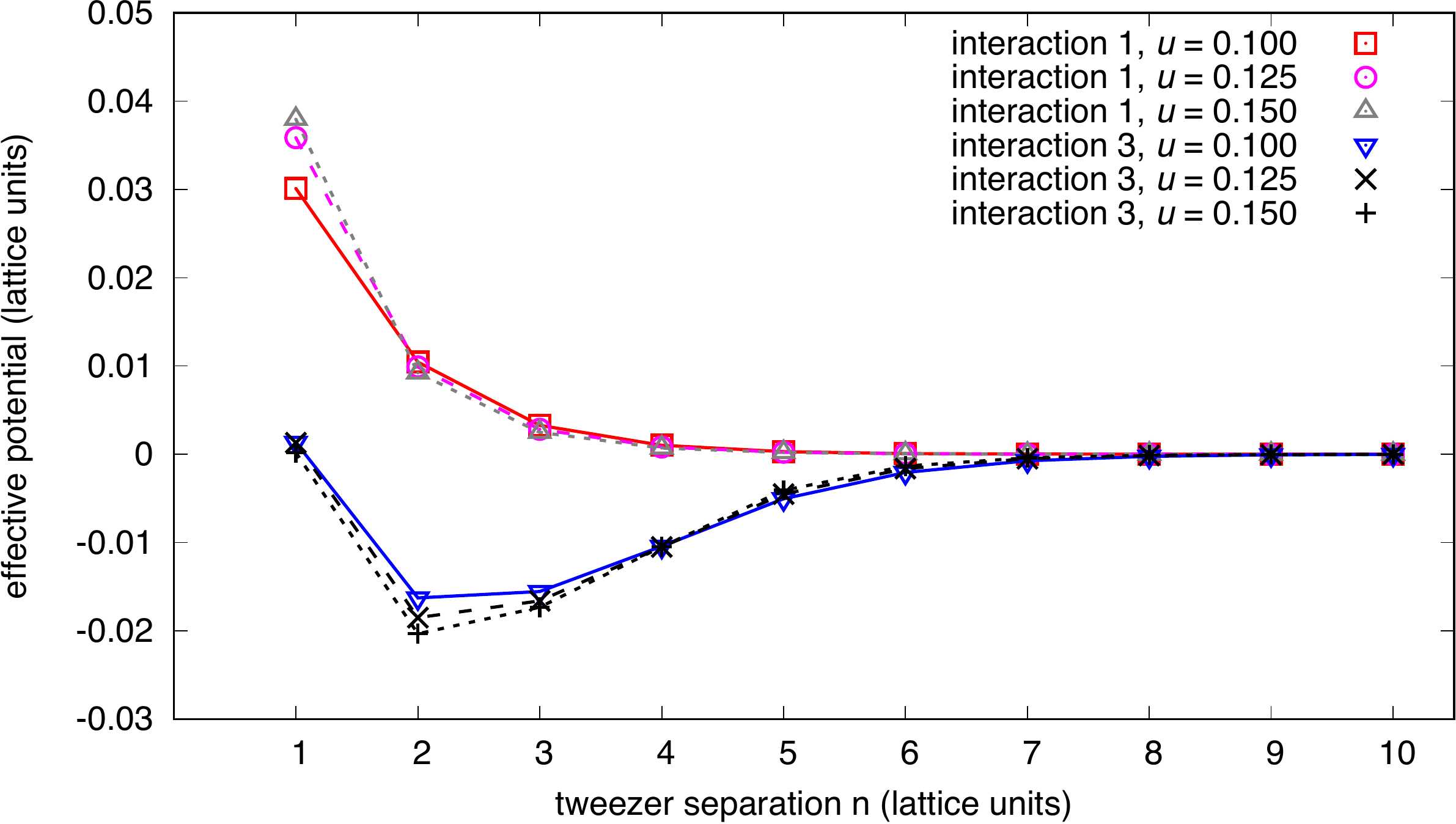}
\label{coefficients}
\end{figure}

In all the results presented here, we
choose the tweezer coupling so that the tweezer binding energy is comparable
to the dimer binding energy, and so the spatial resolution of our probe is comparable
to the dimer size.
Fig.~\ref{coefficients} shows the tweezer effective potential versus separation distance for interactions 1 and 3 and tweezer couplings $u = 0.100, 0.125, 0.150$.  For  $u = 0.100$ the tweezer binding energy is about half the dimer binding energy, while for  $u = 0.150$ it is about the same as the dimer binding energy. We see that while there is some dependence on $u$, a fairly consistent shape can be seen for each interaction.   What is striking is the large difference between the effective potentials.  For  interaction 1 it is purely repulsive while for interaction 3 it is attractive for every separation distance $n>1$.

\begin{figure}[!ht]
\centering
\caption{The tweezer effective potential versus
separation distance for interactions $1,2,3,4,5$ for dimer binding energy
$1/40$ and tweezer coupling $u
= 0.100$.}
\vspace{0.5cm}
\includegraphics[width=8cm]{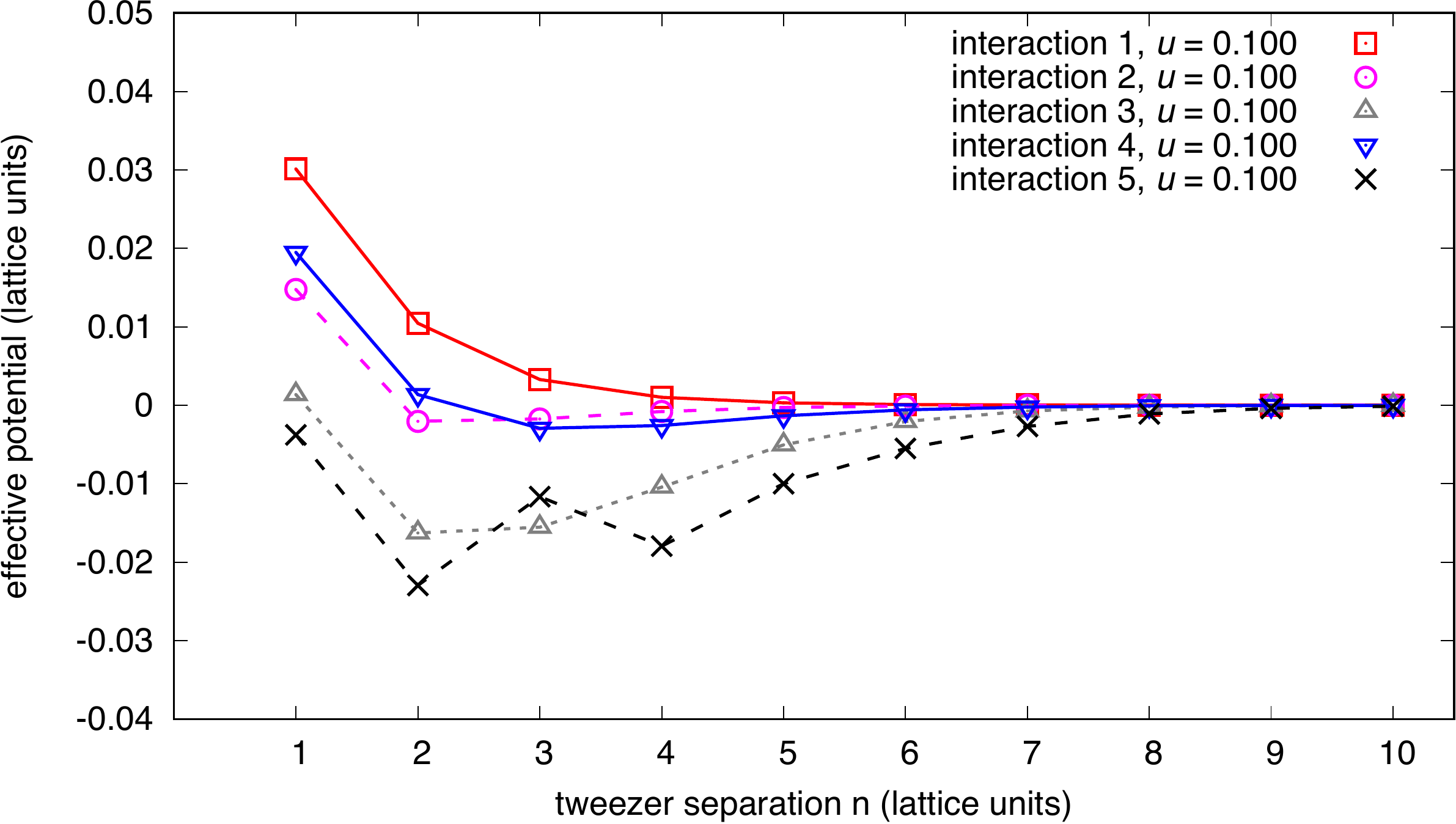}
\label{potentials}
\end{figure}

We now analyze the differences between the interactions more systematically. Fig.~\ref{potentials} shows the tweezer effective potential potential versus
separation distance for tweezer coupling
$u = 0.100$ and interactions $1,2,3,4,$ and $5$. As noted above, the  dimer binding energy for all five interactions has the same value, $1/40$. In all cases we observe that the effective potential becomes more repulsive at short distances, with about the same slope between points $n=1$ and $n=2$ in all five cases.  This suggests that the physics of Pauli repulsion is dominant at short distances.  We note that the tweezer effective potential is not well-defined for $n=0$ due to the Pauli exclusion principle.   

Let us now compare the interactions $1,2,3$. Each of the interactions  $1,2,$ and $3$ are local step potentials with radii equal to $0,1,$ and $2$ lattice units respectively.  We observe that the attraction between dimers increases with the range of the local interaction. As the range increases, the particles are able to interact from longer distances. This makes it easier for opposite spin particles to attract each other while avoiding the strong Pauli repulsion between identical fermions.  We note also that by increasing the range of the local interaction, the attraction in the odd-parity channel increases.  

We now compare the tweezer effective potentials for interactions 3 and 4.  The attraction between the dimers is rather strong for interaction 3 while interaction 4 is mostly repulsive.  By design we have made interactions 3 and 4  identical in the even-parity channel, and so the dimer wave functions are exactly the same. However interaction 4 contains a nonlocal exchange potential that   removes all of the interaction strength in the odd-parity channel.  Therefore one can view the difference between the effective potentials for interactions 3 and 4 as arising from attractive odd-parity interactions in interaction 3 that are completely missing from interaction 4. This is an appealing and simple explanation. We can view the dimers as closed-shell bound states for the even-parity or ``$S$-wave'' interactions, and the attraction between dimers is due to odd-parity or ``$P$-wave''  interactions between closed-shell dimers.  

However this is not the full story.  To illustrate we now turn our attention to the effective potential for interaction 5. Interaction 5 is like interaction 4 in that the nonlocal
exchange potential
removes all the odd-parity interactions.
However the effective potential is even more
attractive for interaction 5 than for interaction 3. How can this be?

\begin{figure}[!ht]
\centering
\caption{Local part of the particle-particle
interaction $I_{i,i}$ versus relative particle separation $i$ for interactions $1,2,3,4,5$ and dimer binding energy $1/40$.}
\vspace{0.5cm}
\includegraphics[width=8cm]{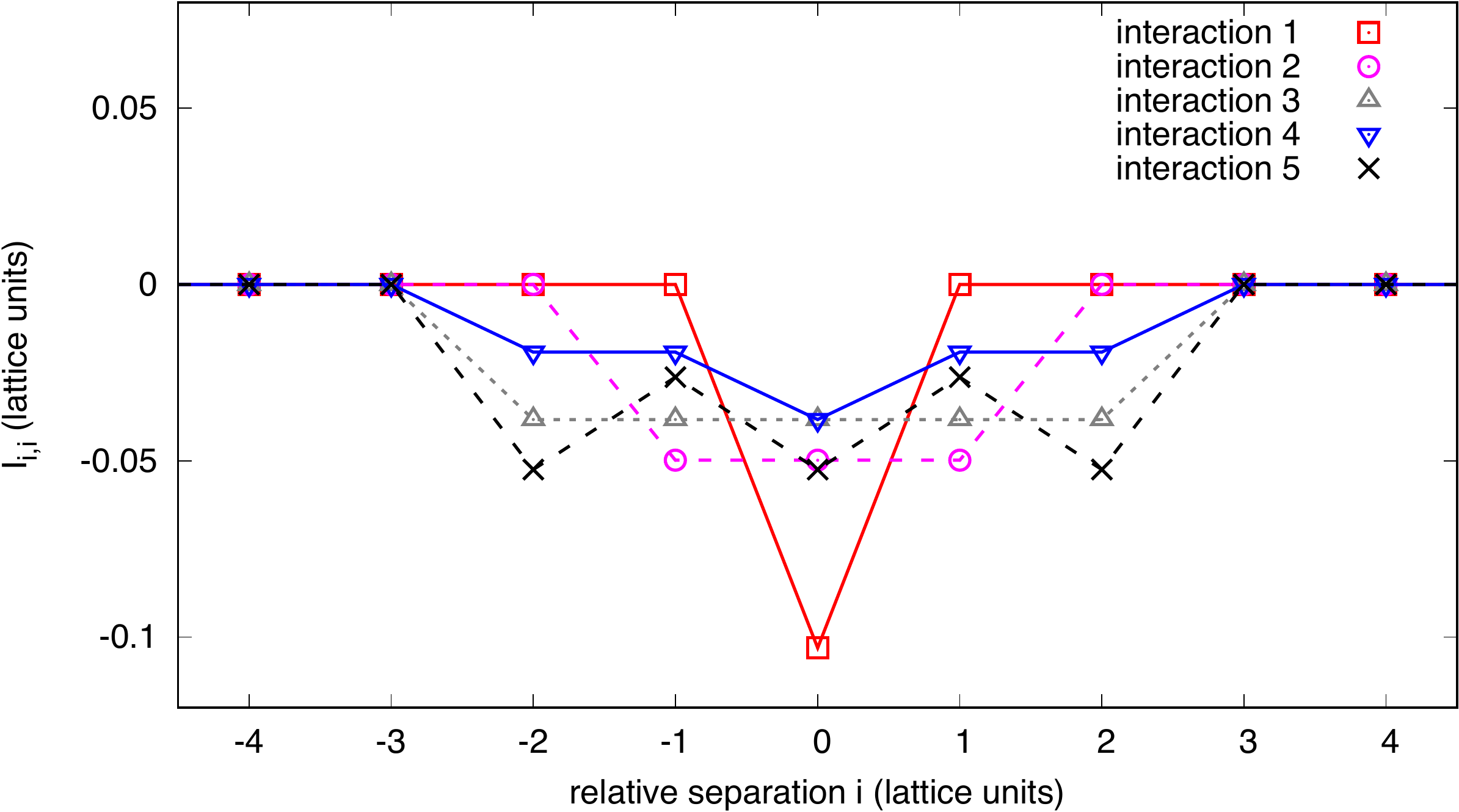}
\label{local}
\end{figure}

An answer to this puzzle emerges when we plot the local part of the particle-particle interaction $I_{i,i}$ versus relative particle separation $i$.  This is  shown in Fig.~\ref{local}.  We see that the local part of interaction 4 is weaker than that of interaction 3, while the local part of interaction 5 is stronger.  This finding is entirely consistent with the conclusions of Ref.~\cite{Elhatisari:2016owd}.  The strength of the local part of the interaction appears to play a dominant role in the effective forces between quantum bound states.

Our earlier explanation of the difference between interactions 3 and 4 being due to attractive odd-parity interactions can be rolled neatly into this notion of local dominance.  We see in
Fig.~\ref{local} that the odd-parity interactions contained in interaction 3 are responsible for enhancing the strength of the local part of the interaction relative to interaction 4.  So one can say that the difference between interactions 3 and 4  is either due to the attractive odd-parity interactions or due to the strength of the local interactions.  Both explanations are equivalent.

The connection between local dominance and attractive higher-partial waves will persist in higher dimensions as well.  In order to show this we make a brief detour and consider a continuum $S$-wave two-body interaction in three dimensions.  The explicit projection onto the $S$-wave produces nonlocality with respect to the incoming and outgoing angles.  The angular integral kernel of the $S$-wave projection operator has the form 
\begin{equation}
Y^*_{0,0}(\theta,\phi)Y_{0,0}(\theta',\phi'),
\end{equation}
where $Y_{L,L_z}$ denote spherical harmonics, $(\theta',\phi')$ are the orientation angles of the incoming relative separation vector between the particles, ${\bf r}'={\bf r}_2'-{\bf r}_1'$, and $(\theta,\phi)$ are the orientation
angles of the outgoing relative separation vector between the particles, ${\bf r}={\bf r}_2-{\bf r}_1$.
When the interaction is sufficiently attractive,
an $S$-wave bound state will form.

We now include an additional attractive interaction in partial wave $L$. For $L>0$ and without partial wave mixing, this will have no impact on the $S$-wave bound state wave function or binding energy. The angular integral kernel of the projection operator onto partial wave $L$ has the
form       
\begin{equation}
\sum_{L_z} Y^*_{L,L_z}(\theta,\phi)Y_{L,L_z}(\theta',\phi').
\end{equation}
We note that the closure relation,
\begin{equation}
\sum_{L_z} Y^*_{L,L_z}(\theta,\phi)Y_{L,L_z}(\theta,\phi) = \frac{2L+1}{4\pi},
\end{equation}
implies that any attractive interaction in partial wave $L$ will increase the attraction in the part of the interaction that is local with respect to angular orientation, $(\theta,\phi)=(\theta',\phi')$.  This shows the direct connection between local dominance and attractive higher-partial waves.
We mention that our arguments above are related to the well-known result that if a local potential is attractive
at every point, it will be attractive in all partial waves.
 
We now return back to the one-dimensional system at hand. A closer look at the tweezer effective potential for interaction 5 shows unusual energy minima at separation distances $n = 2$ and $n = 4$.  To understand how this arises,  we note that the local part of the interaction $I_{i,i}$   has a zigzag shape with energy minima at $i=0$ and $i=\pm 2$.  The locations of these energy minima overlap when the tweezer separations are at $n=2$ or $n=4$, thus creating deep minima where the up-spin particles can simultaneously feel the attractive potential due to both down-spin particles.  The nonlinear response of the up-spin particles to this coherent alignment of the local minima in $I_{i,i}$ gives rise to the energy minima in the effective potential at  $n = 2$ and $n = 4$.

By studying a simple system
of interacting dimers in one spatial dimension we have discovered a rich phenomenology of dimer-dimer interactions and their connection to the underlying particle-particle interactions.  In some cases like interaction 1, we find that the interaction is purely repulsive.  We note that this repulsive interaction between dimers for point-like interactions carries over to the dimer-dimer interaction in three dimensions as well \cite{Petrov:2004zz,Petrov:2005zz,Elhatisari:2016hui}. 
For other cases such as interaction 3 and interaction 5, we get a strongly attractive interaction between dimers.  The strength and range of the local part of the interaction play a dominant role in determining the effective potential.

The reason why local interactions are important was already discussed in Ref. \cite{Elhatisari:2016owd} in the context of the tight-binding approximation.  In the tight-binding limit, where the bound state is a tightly-bound compact object, local interactions keep the interacting particles in their original place and therefore add coherently.  We note that the tight-binding limit with local
interactions
can be viewed as a classical limit
where the bound state wave function sits at the bottom of the potential energy
surface.  In contrast, a general nonlocal interaction will attempt to move particles into locations not interior to the compact bound states.
 In short, the local or nearly local interactions are dominant because they preserve spatial coherence.   

\begin{figure}[!ht]
\centering
\caption{The tweezer effective potential versus
separation distance for interactions $1,2,3,4,5$, with dimer binding energy $1/100$ and tweezer coupling $u
= 0.100$.}
\vspace{0.5cm}
\includegraphics[width=8cm]{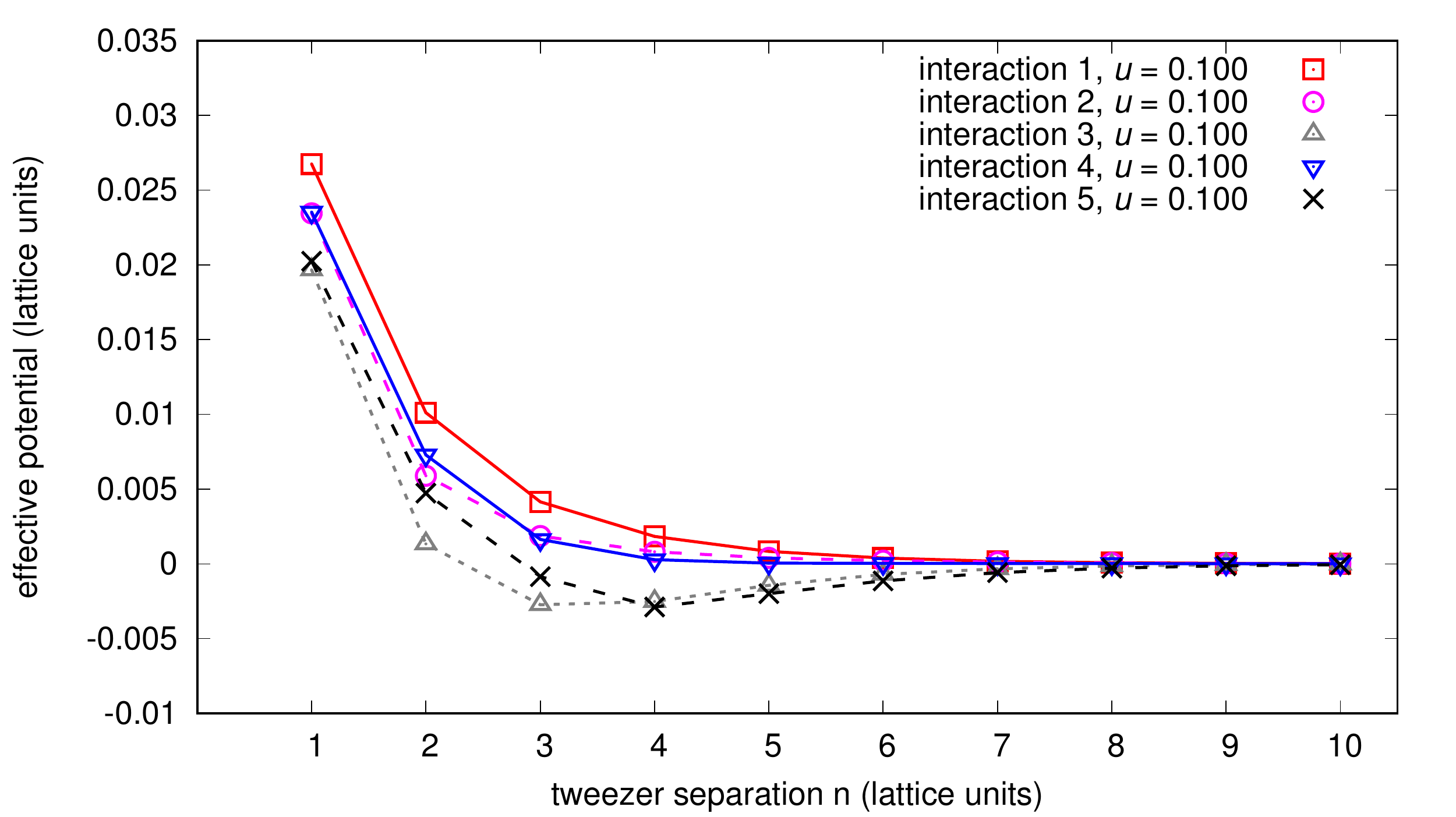}
\label{potentials_1MeV}
\end{figure}

We have seen that decreasing the range of the interaction with the dimer binding energy fixed makes the effective potential more repulsive.  Our zero-range interaction 1 can be viewed as an example of the universal limit when the range of the interaction is much smaller than the dimer size.  We can also see this universal behavior by decreasing the binding energy of the dimers.  In Fig.~\ref{potentials_1MeV} we show the tweezer effective potential versus
separation distance for interactions $1,2,3,4,$ and $5$ with dimer binding energy $1/100$ and tweezer coupling
$u = 0.100$.  The corresponding values of the couplings are $c_1 =-0.064031$,
 $c_2 =-0.027454$, $c_3 =-0.019684$, $c_4 =-0.019684$, and $c_5 =-0.031440$.  We observe that overall the effective potential is less attractive, exhibiting universal behavior when the
range of the interaction is much smaller than the dimer size.

Nevertheless, our analysis of the differences among the five cases is still valid.  When we compare interactions 1, 2, and 3, the effective potential becomes more attractive as the range of the interaction is increased.  The purely even-parity interaction 4 is less attractive than interaction 3, while the purely even-parity interaction 5 is much more attractive than interaction 4 due to the strong local part of the interaction. In the Supplemental Materials we also present an extension of our analysis
to two
spatial dimensions and find results which are similar to the one dimensional results presented here.  

Our conclusions corroborate the findings in Ref. \cite{Elhatisari:2016owd}
and suggest that the phenomenon of local dominance in the effective forces
between quantum bound states is a universal feature appearing in other quantum
systems, regardless of system details and dimensionality.
Indeed, one can understanding the strength of van der Waals, ionic, covalent,
 metallic, and hydrogen bonding as being enhanced by the locality of the
Coulomb interaction.  Our results should also be useful in understanding
the properties of composite bosons in designer quantum systems such as optical
lattices.

There is much work left to do in understanding the nature of effective forces between quantum bound states. Many other systems must be studied with the same depth as we have pursued here. However this simple study of dimers in one spatial dimension has already elucidated many important facets of the general phenomenon.  We hope that these insights will lead to a deeper understanding of the connection between nuclear forces and nuclear structure.  At a more practical level, we are already using some of these findings to develop new lattice chiral interactions with improved order-by-order convergence
in  {\it ab initio} nuclear structure calculations of medium mass nuclei.

\section*{Acknowledgement}
We are grateful for discussions with Lucas
Platter and for the hospitality of the Kavli Institute for Theoretical Physics at UC Santa Barbara for hosting E.E, H.K., and D.L.  We also acknowledge the partial financial support provided by the CRC110: ``DFG (SFB/TR 110, "``Symmetries and the Emergence of 
Structure in QCD''")'', the BMBF (Verbundprojekt 05P2015 - NUSTAR R\&D), and the U.S. Department of Energy (DE-FG02-03ER41260).
This research was also supported in part by the National Science Foundation under Grant No. PHY11-25915.

\clearpage

\renewcommand{\thefigure}{S\arabic{figure}}
\setcounter{figure}{0}

\section*{Supplemental Material}
We generalize our study of the effective potential between dimers to two
dimensions. We consider lattice Hamiltonians for two-component fermions in
two dimensions that have the form\begin{align}
& H=t\sum_{{\bf n},\sigma=\uparrow,\downarrow}
\sum_{{\bf l}=\hat{\bf 1},\hat{\bf 2}}[b_\sigma^{\dagger}({\bf n}+{\bf l})-b_\sigma^{\dagger}({\bf
n})][b_\sigma({\bf n}+{\bf
l})-b_\sigma({\bf n})] \nonumber \\
& + \sum_{{\bf n},{\bf i},{\bf i}'}
b_\uparrow^{\dagger}(\tfrac{{\bf n}+{\bf i}}{2})
b_\downarrow^{\dagger}(\tfrac{{\bf n}-{\bf i}}{2})I_{{\bf i},{\bf i}'}
b_\downarrow(\tfrac{{\bf n}-{\bf i}'}{2})
b_\uparrow(\tfrac{{\bf n}+{\bf i}'}{2}),
\end{align}
where ${\bf n}$, ${\bf i}$, ${\bf i}'$, $\tfrac{{\bf n}\pm{\bf i}}{2}$, and
$\tfrac{{\bf n}\pm{\bf i}'}{2}$ are integer-valued vectors. As in the one-dimensional
case, we take $t=1/20$ in dimensionless lattice units.  In nuclear physics
notation this corresponds with the dimensionless mass parameter $m=10$, where
$t=1/(2m)$. We consider five different interactions, and in the following
we list all the nonzero terms.  In all
five cases the interaction produces a bound dimer with dimensionless energy
$-1/40$.

The first interaction is a point-like interaction with
\begin{equation}
I^{(1)}_{(0,0),(0,0)}=c^{(1)},
\end{equation}
where $c^{(1)}= -0.265354$. The second interaction is a local interaction
that has a range of  one lattice unit,
\begin{equation}
I^{(2)}_{{\bf i},{\bf i}}=c^{(2)},
\end{equation}
for $c^{(2)}= -0.119585$ and integer vectors $|{\bf i}| \le 1$. The third
interaction is a local interaction
that has a range of  two lattice units,\begin{align}
& I^{(3)}_{{\bf i},{\bf i}}=c^{(3)},
\end{align}
for $c^{(3)}=-0.073690$ and integer vectors $|{\bf i}| \le 2$.  The fourth
interaction is a nonlocal interaction that has a range of  two lattice units,
 
\begin{align}
& I^{(4)}_{(0,0),(0,0)}=c^{(4)}, \\
& I^{(4)}_{{\bf i},{\bf i}'}=\frac{c^{(4)}}{4},
\end{align}
for $c^{(4)}=-0.175947$ and integer vectors $|{\bf i}| = 2$ and $|{\bf i}'|
= 2$.  This interaction acts only on the rotationally-invariant sector on
the lattice.     The fifth interaction is another nonlocal interaction with
a range of  two lattice units and acts only on the rotationally-invariant
sector,
\begin{align}
& I^{(5)}_{(0,0),(0,0)}=c^{(5)}, \\
& I^{(5)}_{{\bf i},(0,0)}=I^{(5)}_{(0,0),{\bf i}'}=-\frac{c^{(5)}}{8}, \\
& I^{(5)}_{{\bf i},{\bf i}'}=\frac{c^{(5)}}{4},
\end{align}
where $|{\bf i}| = 2$, $|{\bf i}'| = 2$, and $c^{(5)}=-0.225823.$  In comparison
with interaction 4, the local part of interaction 5 is more strongly attractive.

As in the one-dimensional case,  we introduce an external tweezer potential
of the form
\begin{equation}
U_{\rm ext}({\bf n})=
-u\sum_{n'}b_\downarrow^{\dagger}({\bf n}')b_\downarrow^{\dagger}({\bf n}+{\bf
n}')b_\downarrow({\bf n}+{\bf n}')b_\downarrow({\bf n}'),
\end{equation}
with coefficient $u>0$. We then compute the ground state energy of the full
Hamiltonian with external potential, $H+U_{\rm ext}({\bf n})$, versus separation
distance $n=|{\bf n}|$ where ${\bf n}$ is aligned along one of the coordinate
axes.  

\begin{figure}[!ht]
\centering
\caption{The tweezer effective potential for versus
separation distance for interactions $1,2,3,4,5$, dimer binding energy $1/40$,
and tweezer coupling $u
= 0.170$.}
\vspace{0.5cm}
\includegraphics[width=8cm]{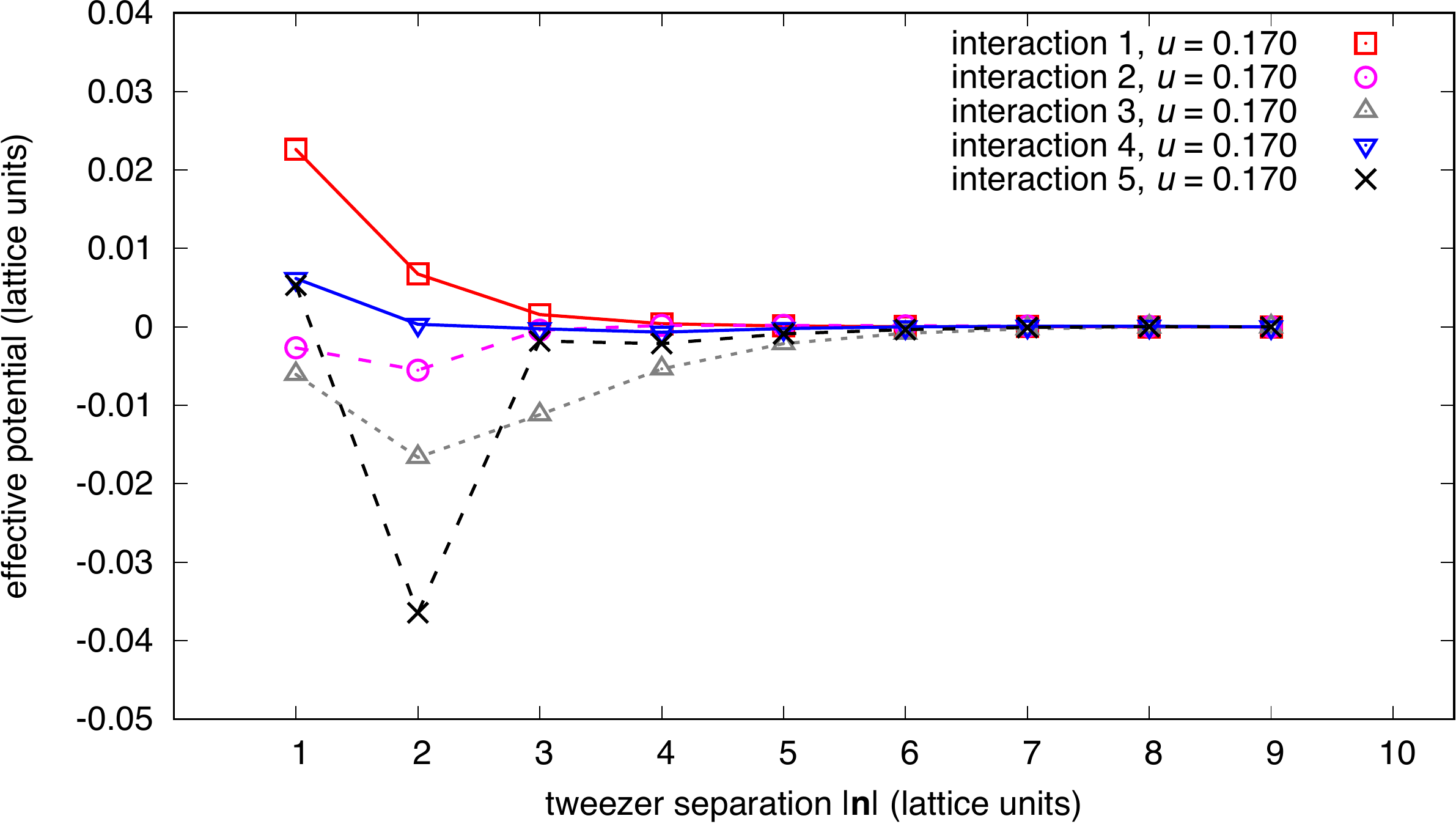}
\label{potentials_2D}
\end{figure}

Fig.~\ref{potentials_2D} shows the tweezer effective potential potential
versus
separation distance for tweezer coupling
$u = 0.170$ and interactions $1,2,3,4,$ and $5$. The  dimer binding energies
for all five interactions have the same value, $1/40$.  Comparing results
for the local step potentials $1,2,$ and $3$, we see that the attraction
between dimers increases with the range of the local interaction.  We also
note that nonlocal interaction 5 is much more attractive than nonlocal interaction
4.  These results are very similar to what was found for the one-dimensional
system.

\begin{figure}[!ht]
\centering
\caption{Local part of the particle-particle
interaction $I_{{\bf i},{\bf i}}$ versus relative particle separation $\pm|{\bf
i}|$ for interactions $1,2,3,4,5$ with dimer binding energy $1/40$.}
\vspace{0.5cm}
\includegraphics[width=8cm]{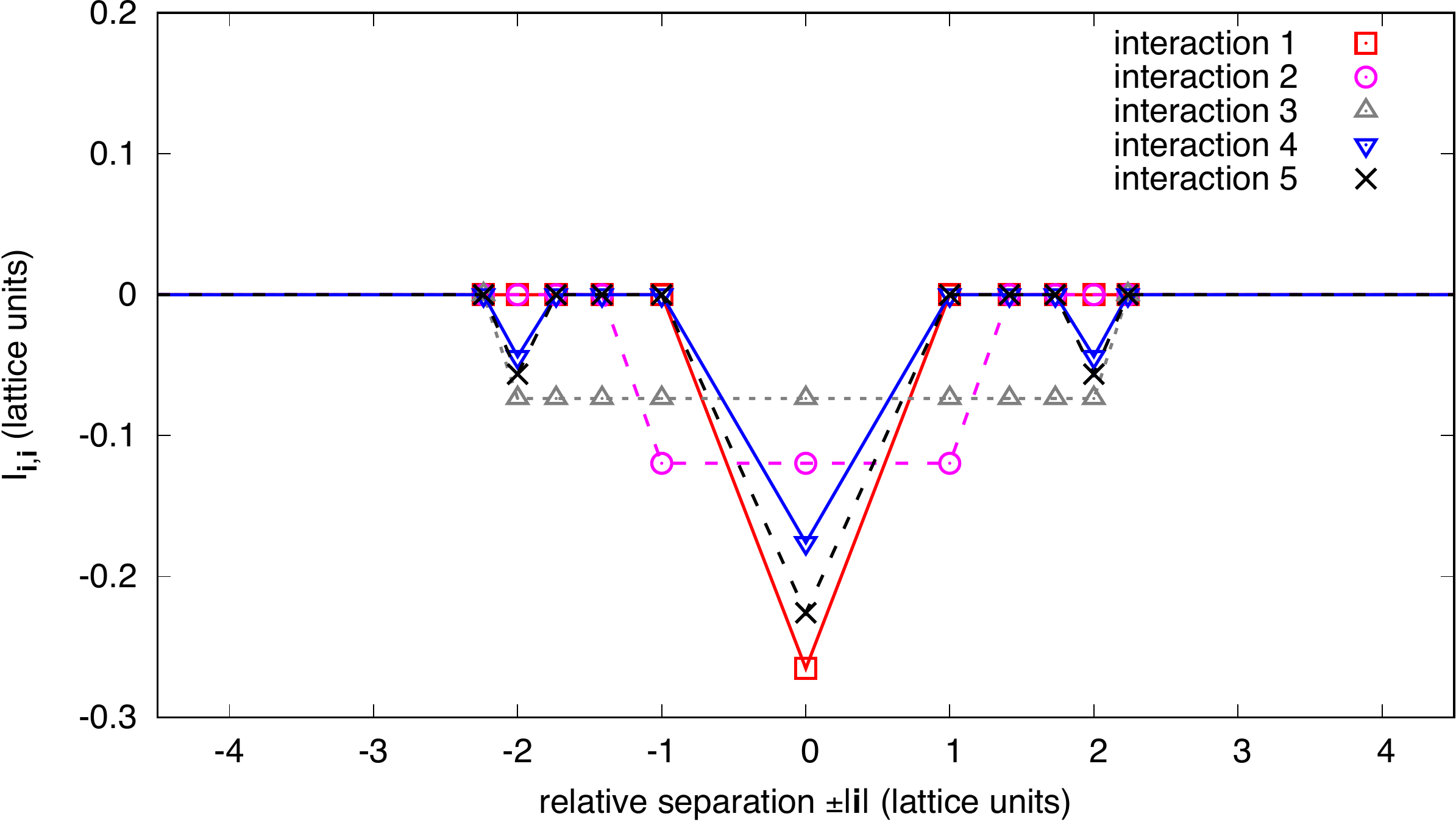}
\label{local_2D}
\end{figure}

In Fig.~\ref{local_2D} we plot the local part of the particle-particle interaction
$I_{{\bf i},{\bf i}}$ versus relative particle separation $\pm |{\bf i}|$.
We see that the local part of
interaction 5 is stronger than that of interaction 4, which explains the
increased attraction in the effective potential.  As in the one-dimensional
case, the local part of the interaction for interaction 5 has a zigzag shape
with energy minima at $|{\bf i}|=0$ and $|{\bf i}|= 2$.  The locations
of these energy minima overlap when the tweezer separations are at $|{\bf
n}|=2$ or $|{\bf n}|=4$, and we note the corresponding minima in the effective
potential.  The effective potential minimum for $|{\bf n}|=4$ is quite shallow
while the minimum for $|{\bf n}|=2$ is rather deep.
 
\begin{thebibliography}{99}
\bibitem{London:1930}
F.~London,
Zeitschrift f{\"u}r Physik {\bf 63}, 245 (1930), doi:10.1007/BF01421741.

\bibitem{Arriola:2007de} 
  E.~Ruiz Arriola,
  arXiv:0709.4134 [nucl-th].
    
\bibitem{Elhatisari:2016owd} 
  S.~Elhatisari {\it et al.},
  Phys.\ Rev.\ Lett.\  {\bf 117}, no. 13, 132501 (2016)
  doi:10.1103/PhysRevLett.117.132501
  [arXiv:1602.04539 [nucl-th]].
  
\bibitem{Maris:2014hga} 
  P.~Maris, J.~P.~Vary, A.~Calci, J.~Langhammer, S.~Binder and R.~Roth,
  Phys.\ Rev.\ C {\bf 90}, no. 1, 014314 (2014)
  doi:10.1103/PhysRevC.90.014314
  [arXiv:1405.1331 [nucl-th]]. 
   
\bibitem{Duguet:2016wwr} 
  T.~Duguet, V.~Som{\`a}, S.~Lecluse, C.~Barbieri and P.~Navrátil,
  arXiv:1611.08570 [nucl-th].
  
\bibitem{Stroberg:2016ung} 
  S.~R.~Stroberg, A.~Calci, H.~Hergert, J.~D.~Holt, S.~K.~Bogner, R.~Roth and A.~Schwenk,
  arXiv:1607.03229 [nucl-th].
    
\bibitem{Ruiz:2016gne} 
  R.~F.~Garcia Ruiz {\it et al.},
  Nature Phys.\  {\bf 12}, 594 (2016)
  doi:10.1038/nphys3645
  [arXiv:1602.07906 [nucl-ex]].  
  
\bibitem{Hagen:2016uwj} 
  G.~Hagen, G.~R.~Jansen and T.~Papenbrock,
  Phys.\ Rev.\ Lett.\  {\bf 117}, no. 17, 172501 (2016)
  doi:10.1103/PhysRevLett.117.172501
  [arXiv:1605.01477 [nucl-th]].  

\bibitem{Dytrych:2016vjy} 
  T.~Dytrych {\it et al.},
  Comput.\ Phys.\ Commun.\  {\bf 207}, 202 (2016)
  doi:10.1016/j.cpc.2016.06.006
  [arXiv:1602.02965 [nucl-th]].  
  
\bibitem{Epelbaum:2008ga} 
  E.~Epelbaum, H.~W.~Hammer and U.-G.~Meissner,
  Rev.\ Mod.\ Phys.\  {\bf 81}, 1773 (2009)
  doi:10.1103/RevModPhys.81.1773
  [arXiv:0811.1338 [nucl-th]].

\bibitem{Combescot:2005}
  M.~Combescot and O.~Betbeder-Matibet,
  Eur. Phys. J. B {\bf 48} 469 (2005), doi:10.1140/epjb/e2006-00007-3.
  
\bibitem{Combescot:2007}
  M.~Combescot, O.~Betbeder-Matibet, and R.~Combescot,
  Phys. Rev. B {\bf 75} 174305 (2007), doi:10.1103/PhysRevB.75.174305.

\bibitem{Combescot:2008}
  M.~Combescot, O.~Betbeder-Matibet, and F.~Dubin,
  Phys. Rep. {\bf 463} 215 (2008), doi:10.1016/j.physrep.2007.11.003.

\bibitem{Combescot:2009}
  W.~V.~Pogosov, M.~Combescot, M.~Crouzeix,
  Phys. Rev. B {\bf 81} 174514 (2010), doi:10.1103/PhysRevB.81.174514. 

\bibitem{Combescot:2010}
  L.~Pilozzi, M.~Combescot, O.~Betbeder-Matibet, A.~D'Andrea,
  Phys. Rev. B {\bf 82} 075327 (2010), doi:10.1103/PhysRevB.82.075327.  
          
\bibitem{Epelbaum:2013paa} 
  E.~Epelbaum, H.~Krebs, T.~A.~L{\"a}hde, D.~Lee, U.-G.~Mei{\ss}ner and G.~Rupak,
  Phys.\ Rev.\ Lett.\  {\bf 112}, no. 10, 102501 (2014)
  doi:10.1103/PhysRevLett.112.102501
  [arXiv:1312.7703 [nucl-th]].
      
\bibitem{Lahde:2013uqa} 
  T.~A.~L{\"a}hde, E.~Epelbaum, H.~Krebs, D.~Lee, U.-G.~Mei{\ss}ner and G.~Rupak,
  Phys.\ Lett.\ B {\bf 732}, 110 (2014)
  doi:10.1016/j.physletb.2014.03.023
  [arXiv:1311.0477 [nucl-th]].
  
\bibitem{Lahde:2015ona} 
  T.~A.~L{\"a}hde, T.~Luu, D.~Lee, U.-G.~Mei{\ss}ner, E.~Epelbaum, H.~Krebs and G.~Rupak,
  Eur.\ Phys.\ J.\ A {\bf 51}, no. 7, 92 (2015)
  doi:10.1140/epja/i2015-15092-1
  [arXiv:1502.06787 [nucl-th]].
  
  
\bibitem{Elhatisari:2015iga} 
  S.~Elhatisari, D.~Lee, G.~Rupak, E.~Epelbaum, H.~Krebs, T.~A.~L{\"a}hde, T.~Luu
and U.-G.~Mei{\ss}ner,
  Nature {\bf 528}, 111 (2015)
  doi:10.1038/nature16067
  [arXiv:1506.03513 [nucl-th]].

\bibitem{Petrov:2004zz} 
  D.~S.~Petrov, C.~Salomon and G.~V.~Shlyapnikov,
  Phys.\ Rev.\ Lett.\  {\bf 93}, 090404 (2004).
  doi:10.1103/PhysRevLett.93.090404

\bibitem{Petrov:2005zz} 
  D.~S.~Petrov, C.~Salomon and G.~V.~Shlyapnikov,
  Phys.\ Rev.\ A {\bf 71}, 012708 (2005)
  doi:10.1103/PhysRevA.71.012708
  [cond-mat/0407579 [cond-mat.stat-mech]].  
        
\bibitem{Elhatisari:2016hui} 
  S.~Elhatisari, K.~Katterjohn, D.~Lee, U.-G.~Mei{\ss}ner and G.~Rupak,
  arXiv:1610.09095 [nucl-th].


\end{thebibliography}
\end{document}